\documentclass[12pt]{amsart}

\pdfoutput = 1

\usepackage{macros}
\usepackage[normalem]{ulem}
\usepackage{float}

\usepackage{tikz-feynman}

\begin{document}
\onehalfspacing
	
\author{Fabian Hahner}

\email{fhahner@uw.edu}

\address{University of Washington, Department of Physics \\ 3910 15th Ave NE, Seattle, WA 98195,  U.S.A}

\title{Moduli spaces of type II twists}

\begin{abstract}
	We provide a complete classification of twisting supercharges for type IIA supergravity in a flat background by determining the orbits of square-zero elements in the $\cN=(1,1)$ super Poincar\'e algebra in ten dimensions under the action of the Spin group. Together with recent results for type IIB, this completes the description of the moduli space of twists for type II supergravities. We show that the type IIA nilpotence variety is the union of the eleven-dimensional nilpotence variety and a second component, confirming a conjecture of~\cite{NV}, and determine which twists of type IIA arise via dimensional reduction. Using the pure spinor formulation of the superstring, we construct maps from the moduli spaces of worldsheet twists to those of target space twists for both type IIA and type IIB, and find that these maps are not surjective. Precisely those twists of supergravity whose stabilizers are special holonomy groups---$G_2$ in type IIA, $\Spin(7)$ and $\Sp(4)$ in type IIB---have no corresponding background on the pure spinor worldsheet. Finally, we work out the action of T-duality relating twists of type IIA and type IIB, and find an obstruction to dualizing the $G_2$-twist of type IIA along certain
	directions.
\end{abstract}
	
	\maketitle
	\thispagestyle{empty}
	
	\setcounter{tocdepth}{1}
	
        \newpage
	\tableofcontents
	
	\setlength{\parskip}{7pt}

\section{Introduction}
Twisted supergravity theories are topological-holomorphic field theories obtained by placing supergravity in a background where the bosonic ghosts for the supersymmetry take a non-vanishing value~\cite{CostelloLi}. They describe protected subsectors of the original theory, often in terms of a moduli problem with a clear geometric interpretation, and are therefore of great interest both for gaining a better understanding of the supergeometric origins of supergravity itself~\cite{RSW, sconf, CY2} and for twisted holography as an exact approach to holographic dualities~\cite{Costello:2018zrm, Costello:2020jbh, Costello:2021kiv, tw-bfss-ikkt}.

Let $\fp = (\fp_+ \oplus \Pi \fp_- , [-,-] )$ be a super Poincar\'e algebra. For a supergravity theory in a flat background $\R^d$, the supersymmetry ghosts are functions $q \in C^\infty(\R^d, \fp_-)$. The equations of motion imply that admissible background values for $q$ have to be covariantly constant and pointwise satisfy $[q, q]=0$. Thus, such backgrounds are described by the space
\begin{equation}
Y = \{ Q \in \fp_- \: | \: [Q,Q] = 0 \}.
\end{equation}
This space, often called the nilpotence variety, is an affine variety and comes with a group action by the simply connected Lie group integrating $\fp_+$ and an action of $\CC^\times$ by rescaling. Points in $Y$ are called twisting supercharges and their orbits classify supergravity twisting backgrounds in flat space. These spaces were systematically studied for super Poincar\'e algebras with up to sixteen supercharges in~\cite{ElliottSafronov, NV}. In~\cite{NV}, some partial results on theories with 32 supercharges were given (see also~\cite{Movshev:2011cy} for an account on the eleven-dimensional case), but several cases of interest remain.

Progress in twisted supergravity has largely come from two directions: First, since the various ten-dimensional supergravity theories are the low energy limits of ten-dimensional string theories, it is natural to expect that topological versions of worldsheet string theory give rise to twisted supergravity theories in target space. In this direction, topological twists of the $\cN=(2,2)$ sigma model give rise to the A and B models~\cite{Witten:1991zz}, which can be viewed as avatars for a twisted worldsheet string theory. This perspective is reflected in the various conjectures on twisted supergravity in ten dimensions~\cite{CostelloLi, CostelloLiTypeI}. Relying on symmetries and dualities, this line of reasoning was also extended to eleven dimensions~\cite{CostelloMtheory2,CostelloMtheory1}.

On the other hand, the definition of twisted supergravity by Costello and Li~\cite{CostelloLi} does not reference string theory; it defines twisted supergravity as a field theory in the BV formalism. Approaching twisted supergravity from that perspective has led to an effort aimed at both providing proofs for the conjectures mentioned above and at developing a deeper understanding of the geometric structures within twisted supergravity independent of a possible string-theoretic origin. In the free limit, twists of ten- and eleven-dimensional supergravity theory were computed in~\cite{spinortwist, MaxTwist}. For eleven-dimensional supergravity, interactions were investigated in~\cite{RSW, CY2}. Recently, type I supergravity was addressed in the language of generalized geometry~\cite{Kupka:2026usx}.

Establishing a direct connection between the worldsheet and target space approaches to twisted supergravity remains challenging, primarily because worldsheet and target space supersymmetry are rarely manifest simultaneously. In the RNS superstring~\cite{Ramond:1971gb, Neveu:1971rx}, for instance, the worldsheet is inherently supersymmetric, whereas target space supersymmetry emerges only a posteriori as a symmetry of the spectrum. Conversely, the GS string makes target space supersymmetry apparent~\cite{Green:1983wt}, while $\kappa$-symmetry ensures the appearance of the correct number of fermionic degrees of freedom. Although this has been given a supergeometric interpretation via the superembedding formalism~\cite{Sorokin:1988jor}, the relation between worldsheet and target space twists remains relatively unexplored.

Here, we provide the classification of odd square-zero elements in the $\cN = (1,1)$ super Poincar\'e algebra in ten dimensions. Together with~\cite{iib-nv}, this gives a complete account of twisting supercharges for type II supergravity theories in ten dimensions. Further, we describe the algebraic geometry of the orbits and show that $Y_{IIA}$ is a union $Y_{IIA} = Y_{11d} \cup C$, where $Y_{11d}$ is the nilpotence variety for the eleven-dimensional super Poincar\'e algebra and $C$ is the product of two pure spinor cones. This confirms a conjecture of~\cite{NV} and shows that the topological twist of type IIA supergravity that cannot be obtained from eleven dimensions. 

We then use these classification results to compare the worldsheet and target space perspectives on twisted supergravity. In the pure spinor superstring formulation, we show that the moduli space of worldsheet twists is a product of two pure spinor cones and describe the map to the the target space nilpotence variety. In particular, we find that this map is not surjective: the twists it misses are precisely those whose stabilizers are special holonomy groups, namely the $G_2$-twist in type IIA and the $\Spin(7)$- and $\Sp(4)$-twists in type IIB. In type IIB, the map also fails to be injective on orbits, since the separation of left- and right-moving modes breaks the R-symmetry that is present in the super Poincar\'e algebra. Further, we identify those twists of supergravity that realize mixed A/B models, finding that they arise either in type IIA or type IIB depending on whether the number of holomorphic directions is even or odd, consistent with the brane spectrum in these theories. Finally, we determine the action of T-duality, which relates the moduli spaces of twists for the type IIA pure spinor superstring with that of type IIB. We show that this map does not descend to a map between the nilpotence varieties $Y_{IIA} \to Y_{IIB}$ with the obstruction arising from the $G_2$-twist of type IIA, which cannot be dualized along its holomorphic direction.


\subsection*{Acknowledgements}
I would like to thank Natalie Paquette, Surya Raghavendran, Ingmar Saberi, and Johannes Walcher for helpful discussions related to this project.  This research is supported by the DOE Early Career Research Program under award DE-SC0022924.

\section{Square-zero elements in type II supersymmetry algebras}

\subsection{Type IIA supersymmetry and pure spinors}
Let $V$ denote the ten-dimensional vector representation of $\Spin(10,\C)$ and $S_\pm$ the two 16-dimensional spinor representations. The symmetric square of these spinor representations decomposes as $\Sym^2(S_\pm) \cong V \oplus Z_\pm$, where $Z_\pm$ are the self-dual and anti-self-dual parts of $\wedge^5 V$. We denote the respective projections on the vector representation by
\begin{equation}
	\gamma_\pm : \Sym^2(S_\pm) \longrightarrow V.
\end{equation}
The $\cN=(1,1)$ supertranslation algebra in ten dimensions is
\begin{equation}
	\ft_{IIA} = \Pi (S_+ \oplus S_-) \oplus V,
\end{equation}
where the only non-vanishing bracket is between two odd elements and is given by $\gamma = \gamma_+ + \gamma_-$. The type IIA super Poincar\'e algebra is obtained by taking the semidirect product $\fp_{IIA} = \mathfrak{so}(V) \ltimes \ft_{IIA}$.

\subsubsection{Pure spinors and spinor orbits}
Recall that the spinor representations can be constructed by choosing a decomposition $V \cong L \oplus L^\vee$, where $L \subset V$ is a maximal isotropic subspace and $L$ and $L^\vee$ are paired under $\langle -,-\rangle_V$. Then the positive and negative chirality spinors are identified with the even and odd exterior powers of $L^\vee$, respectively:
\begin{equation}
S_+ = \wedge^{\mathrm{even}} L^\vee = \wedge^0 L^\vee\oplus \wedge^2 L^\vee \oplus \wedge^4 L^\vee  \qquad \text{and} \qquad S_- = \wedge^{\mathrm{odd}} L^\vee = \wedge^1 L^\vee \oplus \wedge^3 L^\vee \oplus \wedge^5 L^\vee.
\end{equation}
We pick a top form $\Omega \in \wedge^5 L^\vee$ with corresponding inverse polyvector $\Omega^{-1} \in \wedge^5 L$. Expanding a general element $\psi \in S_\pm$ in components of different form degrees, we can express the maps $\gamma_\pm$ through wedge products and contractions. 
\begin{equation}
\begin{split}
\gamma_+(\psi, \psi) &= \Omega^{-1}(\psi^{(0)} \wedge \psi^{(4)}) - \frac{1}{2} \Omega^{-1}(\psi^{(2)} \wedge \psi^{(2)}) + \psi^{(2)} \vee \Omega^{-1}(\psi^{(4)}) \\
\gamma_-(\psi, \psi) &= \Omega^{-1}(\psi^{(1)} \wedge \psi^{(3)}) - \frac{1}{2}\Omega^{-1}(\psi^{(3)}) \vee \psi^{(3)} + \psi^{(1)} \wedge \Omega^{-1}(\psi^{(5)})
\end{split}
\end{equation}
To every $\psi \in S_\pm$, we can associate an isotropic subspace given by its annihilator under Clifford multiplication:
\begin{equation}
\mathrm{Ann}(\psi) = \{ v \in V \: | \: v \cdot \psi = 0 \} \subset V.
\end{equation}
Recall that $\psi$ is called pure if $\mathrm{Ann}(\psi)$ is maximally isotropic. In this case, we will use the notation $\mathrm{Ann}(\psi)=: L_\psi$. Conversely, every maximal isotropic subspace of $V$ defines a pure spinor up to scale. Thus, the projective variety of pure spinors is given by orthogonal Grassmannians
\begin{equation}
	\cP_\pm = \mathrm{OG}(5,10)_\pm \subset \P(S_\pm).
\end{equation}
Affine versions are obtained as cones over the orthogonal Grassmannians and will be denoted $\cC \cP_\pm$. Crucially, a spinor $\psi \in S_\pm$ is pure if and only if $\gamma_\pm(\psi, \psi)=0$.

The spin representations decompose into orbits under the action of $\Spin(10)$. Note that we can decompose the Lie algebra as
\begin{equation}
\mathfrak{so}(10) \cong \wedge^2 L \oplus \mathfrak{gl}(L) \oplus \wedge^2 L^\vee .
\end{equation}
In this identification, these summands act on $S_\pm$ with form degree $-2$, $0$, and $2$ with $\wedge^2 L$ acting by contraction and $\wedge^2 L^\vee$ by wedge product. The orbit decomposition of spin representations was studied by Igusa~\cite{Igusa}. In ten dimensions, we find for $S_\pm$ that there are two non-trivial orbits.  
\begin{itemize}
	\item[---] \emph{The pure spinor orbit:} Here, the stabilizer of a point can be identified with $\mathrm{SL}(L) \ltimes \wedge^2 L$.
	\item[---] \emph{The generic (open) orbit:} Here, the stabilizer of a point is isomorphic to $\Spin(7) \ltimes \CC^8$, where $\CC^8$ transforms in the spinor representation of $\Spin(7)$.
\end{itemize}
Most of the square-zero elements both in the type IIA and IIB supersymmetry algebras can be constructed from a pair of pure spinors. We recall the following facts that are useful for our classification purposes (for more details, see~\cite{iib-nv}).
\begin{lem} \label{lem: ps}
	Let $(\psi_1, \psi_2)$ be a pair of pure spinors associated to maximally isotropic subspaces $L_{\psi_1} , L_{\psi_2} \subset V$. Let $r = \dim(L_{\psi_1} \cap L_{\psi_2})$ be the intersection dimension of a pair of pure spinors.
	\begin{itemize}
		\item[1.] $\psi_1$ and $\psi_2$ are of the same chirality if $r$ is odd and of opposite chirality if $r$ is even.
		\item[2.] Let $(\psi'_1 , \psi'_2)$ be another pair of pure spinors with intersection dimension $r'$. They are related to $(\psi_1, \psi_2)$ up to scale via the diagonal action of $\Spin(V)$ if and only if $r' =r$.
	\end{itemize}
\end{lem}
If we view $V = V_{\R} \otimes_\RR \C$ as the complexification of a real vector space $V_{\R}$ equipped with an inner product of Euclidean signature, then choosing a pair $(\psi_1, \psi_2)$ of pure spinors is equivalent to choosing a pair of complex structures for $V_{\R}$. Together, they equip $V_{\R}$ with a transverse holomorphic foliation (THF structure) that identifies $V_{\R} \cong \R^{10-2r} \times \C^r$.

\subsection{Classification of twisting supercharges for type IIA supersymmetry}
\begin{figure}
	\begin{tikzcd}[row sep=2.5em, column sep=0.5em, every node/.style={inner sep=3pt}]
	& {\boxed{\begin{gathered}22\\[-0.7em] {\scriptstyle G_2 \: , \: \CC\times\RR^8}\end{gathered}}} 
	& & {\boxed{\begin{gathered}21\\[-0.7em] {\scriptstyle \mathrm{SL}(5) \: , \: \RR^{10}}\end{gathered}}} 
	& \\
	& & {\boxed{\begin{gathered}20\\[-0.7em] {\scriptstyle \mathrm{SL}(2) \times \mathrm{SL}(3) \: , \: \CC^2\times\RR^6}\end{gathered}}} 
	\arrow[ul] \arrow[ur] 
	& & \\
	& & {\boxed{\begin{gathered}15\\[-0.7em] {\scriptstyle \mathrm{SL}(4) \: , \: \CC^4\times\RR^2}\end{gathered}}} 
	\arrow[u] 
	& & \\
	& {\boxed{\begin{gathered}10\\[-0.7em] {\scriptstyle \mathrm{SL}(5) \: , \: \CC^5}\end{gathered}}} 
	\arrow[ur] 
	& & {\boxed{\begin{gathered}10\\[-0.7em] {\scriptstyle\mathrm{SL}(5) \: , \: \CC^5}\end{gathered}}} 
	\arrow[ul] 
	& \\
	& & * \arrow[ul] \arrow[ur] 
	& & 
	\end{tikzcd}
	\caption{Orbit stratification of the nilpotence variety in IIA.}\label{eq:strat-iia}
\end{figure}
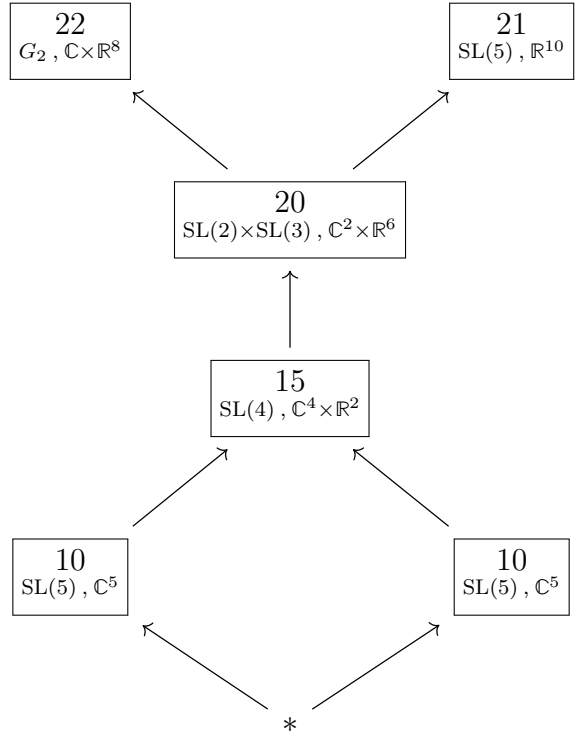

We expand a general supercharge in the type IIA supertranslation algebra as
\begin{equation}
	Q = \psi_+ + \psi_-
\end{equation}
with $\psi_\pm \in S_\pm$ two spinors of opposite chirality. In the following, we discuss solutions to the equation $[Q,Q]=0$ by distinguishing between three cases: 
\begin{itemize}
	\item[---] We can restrict to $\psi_+ = 0$ or $\psi_- = 0$. Then, the non-vanishing component has to be a pure spinor and we obtain holomorphic twisting supercharges.
	\item[---] Both $\psi_+$ and $\psi_-$ are non-vanishing and pure spinors. Then, we obtain different mixed topological-holomorphic supercharges.
	\item[---] Both $\psi_+$ and $\psi_-$ are impure spinors but arranged in such a way that $\gamma(\psi_+, \psi_+) = - \gamma(\psi_- , \psi_-)$.
\end{itemize}
We now describe how these cases decompose into orbits of $\Spin(10)$.
The results on the classification and properties of orbits are summarized in Table~\ref{table-iia}; the stratification of the variety is in Figure~\ref{eq:strat-iia}.

\numpar[][The holomorphic twists]
Holomorphic twists are obtained by choosing a single pure spinor as a twisting supercharge. Since $\Spin(10,\C)$ acts transitively on the space of pure spinors, there are precisely two such choices with $Q$ either being a pure spinor of positive or negative chirality. Thus, we find two distinct orbits isomorphic to the spaces of pure spinors in $S_+$ or $S_-$. These twists are holomorphic and their stabilizers are isomorphic to $\mathrm{SL}(5) \ltimes \wedge^2 L^\vee$.
The twisted theories can naturally be defined on general Calabi--Yau five-folds and depend holomorphically on them.

\numpar[][Mixed type twists from pairs of pure spinors]
Let $\psi_+$ and $\psi_-$ denote two non-vanishing pure spinors and let $L_{\psi_\pm}$ denote the associated maximal isotropic subspaces. Then, the dimension of the intersection $L_{\psi_+} \cap L_{\psi_-}$ is even and gives a complete invariant of square-zero elements constructed by Lemma~\ref{lem: ps}. Further, $r = \dim(L_{\psi_+} \cap L_{\psi_-})$ is the number of surviving holomorphic directions in the twist. 

Thus, we find three distinct orbits of twisting supercharges corresponding to $r \in \{4,2,0\}$. In the following, when we describe representatives for these orbits, we fix a basis $(e_1, \dots e_5)$ of $L$ with dual basis $(e_1^\vee, \dots e_5^\vee)$ and further, by identifying $L_{\psi_+} = L$, we set $\psi_+=1$.

For $r=4$, we can take 
\begin{equation}
	Q = 1 + e_5^\vee
\end{equation}
as a representative. It is straightforward to compute the stabilizer: Decompose
\begin{equation}
	g = (X_- , A, X_+) \in \mathfrak{so}(10)
\end{equation}
so that $X_- \in \wedge^2 L$ acts via contraction, $A \in \mathfrak{gl}(L)$ as a degree-preserving endomorphism, and $X_+ \in \wedge^2 L^\vee$ via wedge product. Then, we see that for $Q$ to be invariant, we need $X_+=0$, while $X_-$ can be arbitrary. Further, we find a block diagonal decomposition for $A$ with the diagonal $4\times 4$-block contributing as $\mathfrak{sl}(4)$ to the stabilizer. The upper-triangular block has to vanish while the lower-triangular block gives four additional generators that transform in the fundamental representation of $\mathfrak{sl}(4)$ and combine with $X_-$ to a 14-dimensional nilpotent component. After integrating to the corresponding Lie group, we thus find for the stabilizer $\mathrm{SL}(4) \ltimes N_{14}$. The twists have 2 topological directions and can be placed on backgrounds of the form $X_4 \times Y_1$ where $X_4$ is a Calabi--Yau 4-fold on which the theory depends holomorphically, while $Y_1$ is a 1-fold on which the theory depends topologically.

For $r=2$, a representative is given by
\begin{equation}
	Q = 1 + e_3^\vee \wedge e_4^\vee \wedge e_5^\vee.
\end{equation}
The computation of the stabilizer proceeds in the same way as for $r=4$. One finds $(\mathrm{SL}(3) \times \mathrm{SL}(2)) \ltimes N_{13}$. The twist has six topological directions so that it can be placed on products $X_3 \times Y_2$ with holomorphic dependence on the Calabi--Yau 3-fold $X_3$ and topological dependence on the 2-fold $Y_2$.

For $r=0$, we can choose $Q = 1 + e_1^\vee \wedge \dots  \wedge e_5^\vee$ as a representative. The stabilizer is $\mathrm{SL}(5)$. The twist is topological, but the stabilizer still requires the background geometry to be a Calabi--Yau 5-fold.

\numpar[][The impure spinor orbit]
A special orbit of twisting supercharges can be constructed by taking both $\psi_+$ and $\psi_-$ to be from the impure spinor orbit with the additional requirement that
\begin{equation}
	\gamma_+(\psi_+, \psi_+) = - \gamma_-(\psi_- , \psi_-) = f.
\end{equation}
The stabilizer of this twisting supercharge is the joint stabilizer of $\psi_+$ and $\psi_-$ inside $\Spin(10, \C)$. Note that $f \in V$ is a null vector and we can expand
\begin{equation}
	V = \textbf{8}_v \oplus \langle f \rangle \oplus \langle f^\vee \rangle,
\end{equation}
where we have chosen $f^\vee$ such that $\langle f, f^\vee \rangle_V = 1$.
Here, and in the following, we will denote the three irreducible eight-dimensional representations of $\Spin(8)$ by $\textbf{8}_v$, $\textbf{8}_s$, and $\textbf{8}_c$. The stabilizer of $f$ inside $\Spin(10)$ is the parabolic subgroup $\Spin(8,\C) \ltimes \textbf{8}_v \otimes \langle f \rangle$.

Let us now decompose the spin representations $S_\pm$ under $\Spin(8)$. Splitting the maximal isotropic in the ten-dimensional vector representation as $L = L_4 \oplus \langle f \rangle \subset V$, we get
\begin{equation}
\begin{split}
	S_+ &= \left( \wedge^0 L_4^\vee \oplus \wedge^2 L_4^\vee \oplus \wedge^4 L_4^\vee \right) \oplus \left( \wedge^1 L_4^\vee \oplus \wedge^3 L_4^\vee \right) \otimes \langle f^\vee \rangle \cong \textbf{8}_s \oplus \textbf{8}_c \otimes \langle f^\vee \rangle \\
	S_- &= \left( \wedge^1 L_4^\vee \oplus \wedge^3 L_4^\vee \right) \oplus \left( \wedge^0 L_4^\vee \oplus \wedge^2 L_4^\vee \oplus \wedge^4 L_4^\vee \right) \otimes \langle f^\vee \rangle \cong \textbf{8}_c \oplus \textbf{8}_s \otimes \langle f^\vee \rangle.
\end{split}
\end{equation}
Note that $f \cdot \psi_\pm =0$, where $\cdot$ denotes Clifford multiplication. Using the above decomposition, this implies $\psi_+ \in \textbf{8}_s$ and $\psi_- \in \textbf{8}_c$.

By triality, there are three different embeddings of $\Spin(7)$ into $\Spin(8)$ fixing either a positive chirality spinor, a negative chirality spinor, or a vector. We take the embedding that keeps $\psi_+ \in \textbf{8}_s$ fixed; then $\textbf{8}_c$ becomes the irreducible spinor representation of $\Spin(7)$. The stabilizer of $\psi_-$ inside $\Spin(7)$ is $G_2$.

Further, we note that the entire unipotent part $\textbf{8}_v \otimes \langle f \rangle$ stabilizes both $\psi_+$ and $\psi_-$. It decomposes into a seven dimensional and a trivial representation of $G_2$ so that the total stabilizer of a supercharge of this type is $G_2 \ltimes (V_7 \oplus 1_{G_2})$.

A representative for such a twist is given by $1 + e_{2345}^\vee + e_2^\vee - e_{345}^\vee$. Then, we have $f=e_1$ and a short computation shows that
\begin{equation}
	\Im[Q,-] = \langle e_1, \dots , e_5 , e_2^\vee , \dots , e_5^\vee \rangle \subset V.
\end{equation}
Thus, the twist has a single surviving holomorphic translation and we expect that such twists can be placed on manifolds $M_8 \times \Sigma$, where $\Sigma$ is a complex manifold of dimension one and $M_8$ is a product between a one-dimensional real manifold and a $G_2$-manifold.

\numpar[][Summary]
We summarize the list of orbits in the variety of square-zero elements in the type IIA supersymmetry algebra in Table~\ref{table-iia}.

\begin{table}[h]
	\centering
	\renewcommand{\arraystretch}{1.5}
	\begin{tabular}{|l|l|l|l|}
		\hline
		Orbit & Stabilizer & Background & Representative \\ \hline \hline
		\underline{Single Pure:} & & & \\
		pure, $+$ & $\mathrm{SL}(5) \ltimes \Lambda^2 L^\vee$ & $\mathbb{C}^5$ & $1$ \\
		pure, $-$ & $\mathrm{SL}(5) \ltimes \Lambda^2 L^\vee$ & $\mathbb{C}^5$ & $e_{12345}^\vee$ \\ \hline
		\underline{Pair Pure:} & & & \\
		$r=4$ & $\mathrm{SL}(4) \ltimes N_{14}$ & $\mathbb{C}^4 \times \mathbb{R}^2$ & $1 + e_5^\vee$ \\
		$r=2$ & $(\mathrm{SL}(2) \times \mathrm{SL}(3)) \ltimes N_{13}$ & $\mathbb{C}^2 \times \mathbb{R}^6$ & $1 + e_{345}^\vee$ \\
		$r=0$ & $\mathrm{SL}(5)$ & $\mathbb{R}^{10}$ & $1 + e_{12345}^\vee$ \\ \hline
		\underline{Impure} & & & \\
		& $G_2 \ltimes (V_7 \oplus 1_{G_2})$ & $\mathbb{C} \times \mathbb{R}^8$ & $1 + e_{2345}^\vee + e_2^\vee - e_{345}^\vee$ \\ \hline
	\end{tabular} 
	\vspace{0.4cm}
	\caption{Orbits of twisting supercharges in IIA.} \label{table-iia}
\end{table}
The representatives are with respect to a basis $(e_1^\vee, \dots, e_5^\vee)$ of $L^\vee$ and we use the shorthand notation $e_{ij}^\vee = e_i^\vee \wedge e_j^\vee$ etc. $N_d$ denotes a nilpotent Lie group of dimension $d$.

While we work with the complexified super Poincar\'e algebra to classify square-zero supercharges, the supergravity theories that are to be twisted are formulated on real spacetime manifolds. To make this connection, we view the vector representation $V$ as the complexification of the real vector representation of $V = V_{\RR} \otimes_{\RR} \CC$ where $V_{\R}$ is equipped with the inner product of Euclidean signature.

The choice of a square-zero supercharge equips $V_{\R}$ with a subspace defining a transverse holomorphic structure that identifies $V_{\R} \cong \R^{10-2r} \times \C^r$ for some $r = 0, \dots 5$. This structure is displayed in the `background' column in the table; the twisted theory will depend topologically on the real and holomorphically on the complexified directions. In general, the twisted theory can be formulated on manifolds equipped with a transverse holomorphic foliation of the appropriate type. Accordingly, the relevant structure groups for these real geometries are the compact real forms of the complex stabilizers listed above. For instance, when discussing the holonomy of curved backgrounds, we will typically refer to the unitary groups $\mathrm{SU}(n)$ rather than the complex special linear groups $\mathrm{SL}(n, \C)$.

The variety of square-zero elements admits the stratification described by Figure~\ref{eq:strat-iia}. Here, we label orbits by their projective dimensions together with the Levi part of their stabilizers and the expected background dependence for the twisted supergravity theory.

Some of the strata have a straightforward explicit description. The two ten-dimensional strata are given by the variety of pure spinors, i.e. they are identified with the orthogonal Grassmannians $OG(5,10)_\pm$, respectively. The 21-dimensional stratum that is the closure of the orbit of topological twists is identified with the product of the two pure spinor cones modulo global rescalings $(\cC \cP_+ \times \cC \cP_-)/\CC^\times$. The global structure of the orbits with $r=2$ and $r=4$ are more complicated to describe. However, they can both be understood as bundles over $\mathrm{OG}(5,10)_+$ in the following way.

For $r=4$, we first choose a pure spinor $\psi_+$, which corresponds to a point in $\mathrm{OG}(5,10)_+$. Next, we have to choose a four-dimensional subspace of $L_{\psi_+}$ and extend it to a maximal isotropic subspace assigned to a negative chirality spinor $\psi_-$. The first part means choosing a point in $\mathrm{Gr}(4,5) \cong \P^4$; the extension to $L_{\psi_-}$ is unique given the requirement that $\psi_-$ is of negative chirality. Finally, there is a factor of $\CC^\times$ given by the relative scale between the two spinors. In total, the orbit therefore is locally a product of the form $\mathrm{OG}(5,10)_+ \times \P^4 \times \C^\times$.

For $r=2$, we have to choose a two-dimensional subspace $K \subset L_{\psi_+}$ which means choosing a point in $\mathrm{Gr}(2,5)$. The extension of this space to $L_{\psi_-}$ is, in this case, not unique. Indeed, choosing such an extension means choosing a maximal isotropic in $K^\perp/K$, which is a point in $\mathrm{OG}(3,6)$. With the relative scale between $\psi_+$ and $\psi_-$, we thus find that the local structure of the orbit is $\mathrm{OG}(5,10) \times \mathrm{Gr}(2,5) \times \mathrm{OG}(3,6) \times \C^\times$.

Finally, the 22-dimensional stratum is identical to the nilpotence variety for eleven-dimensional supergravity whose geometry was studied in~\cite{Movshev:2011cy}.

\subsection{Review of type IIB}
For completeness and for reference to~\S\ref{sec: app}, let us briefly summarize the orbit stratification of the space of square-zero elements in the type (2,0) super Poincar\'e algebra. These results were published in~\cite{iib-nv}.

In order to define the IIB supersymmetry algebra, we fix a two-dimensional vector space equipped with a non-degenerate symmetric inner product $W=(\CC^2, (-,-)_W)$. Then the complex $\cN=(2,0)$ supertranslation algebra in ten dimensions is the super Lie algebra with underlying $\ZZ_2$-graded vector space
\begin{equation}
\ft_{IIB} =  \Pi (S_+ \otimes W) \oplus V ,
\end{equation}
where the only non-vanishing brackets appear between two odd elements and are given by $\gamma \otimes (-,-)_W$. We extend this super Lie algebra to the ten-dimensional $\cN=(2,0)$ super Poincar\'e algebra by taking the semidirect product $\fp_{IIB} = \left( \mathfrak{so}(V) \oplus \mathfrak{o}(W) \right) \ltimes \ft_{IIB}$.
\begin{table}[H]
	\centering
	\begin{tabular}{|l|l|l|l|l|}
		\hline
		& & & \\[-0.3cm]
		\text{Orbit} & \text{Stabilizer} & \text{Background} & \text{Representative} \\ \hline \hline
		& & & \\[-0.3cm]
		\underline{Rank 1:} & & & \\
		& & & \\[-0.3cm]
		$\{p\}$, iso & $\mathrm{SL}(5) \ltimes N_{10} \times \CC^\times$ & $\CC^5$ & $1 \otimes (1,i)$ \\ 
		& & & \\[-0.3cm]
		$\{p\}$, non-iso & $\mathrm{SL}(5) \ltimes N_{10}$  & $\CC^5$ & $1 \otimes (1,0)$\\ 
		& & & \\[-0.3cm]
		$\emptyset$, iso & $\Spin(7) \ltimes N_8 \times \CC^\times$ & $\R^{8} \times \CC$ & $(e_{23}^\vee + e_{45}^\vee) \otimes (1,i)$ \\ 
		& & & \\[-0.3cm] \hline
		& & & \\[-0.3cm]
		\underline{Rank 2:} & & & \\
		& & & \\[-0.3cm]
		$\mathrm{line}$ & $(\mathrm{SL}(2) \times \mathrm{SL}(3)) \ltimes N_{15} \times \CC^\times$ & $\RR^4 \times \CC^3$ & $1 \otimes (1,0) + e^\vee_{45} \otimes (0,1)$\\
		& & & \\[-0.3cm]
		$\{p_1,p_2\}$ & $\mathrm{SL}(4) \ltimes N_8$ & $\RR^8 \times \CC$ & $1 \otimes (1,0) + e^\vee_{2345} \otimes (0,1)$\\
		& & & \\[-0.3cm]
		$\{p\}$ & $\mathrm{Sp}(4) \ltimes N_{13} \times \CC^\times$& $\RR^{8} \times \CC$ & $1 \otimes (1,0) + (e^\vee_{23} + e^\vee_{45}) \otimes (1,i)$\\
		& & &\\[-0.3cm] \hline
	\end{tabular}
	\vspace{0.4cm}
	\caption{Orbits of twisting supercharges for type IIB.}
\end{table}

In this table, the orbits are labeled as follows. We view twisting supercharges as maps $Q: W^\vee \longrightarrow S_+$ and then distinguish first by the rank of this map and then by the intersection pattern of its projectivized image with the pure spinor variety (where the possibilities are the empty set, one or two points, and a line). In the rank one case, twisting supercharges of the form $\psi \otimes w \in S_+ \otimes W$ with $\psi$ a pure spinor are further distinguished by whether $w$ is isotropic or not. For more information, we refer the reader to~\cite{iib-nv}.

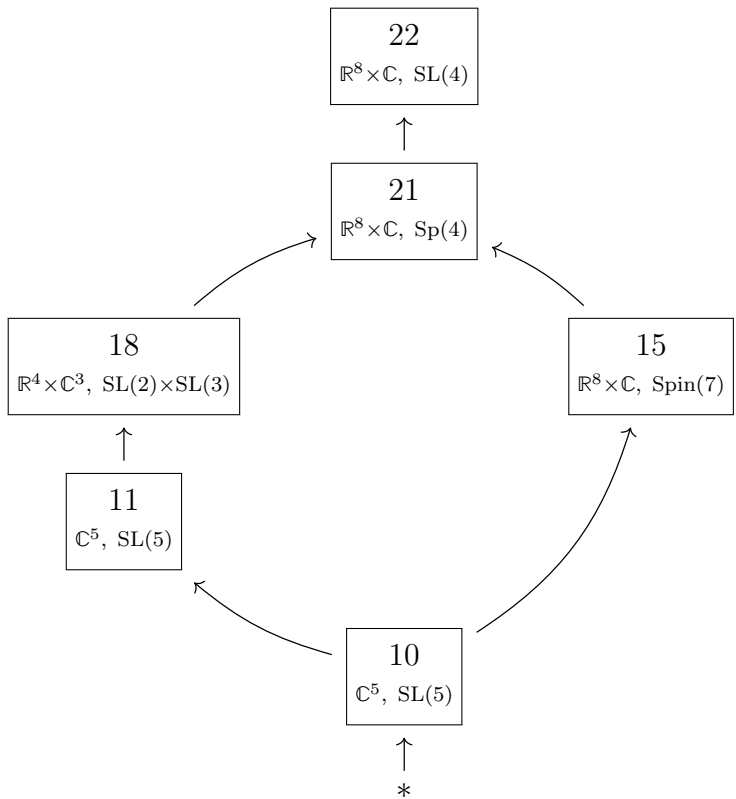
\begin{figure}[H]
	\begin{tikzcd}[row sep=1.1em, column sep=2em, every node/.style={inner sep=3pt}]
	& & & {\boxed{\begin{gathered}22\\[-0.7em] {\scriptstyle{\RR^8\times\CC,\ \mathrm{SL}(4)} }\end{gathered}}} & & & & \\
	& & & {\boxed{\begin{gathered}21\\[-0.7em] {\scriptstyle \RR^8\times\CC,\ \mathrm{Sp}(4)}\end{gathered}}} 
	\arrow[u] & & & & \\
	& &  {\boxed{\begin{gathered}18\\[-0.7em] {\scriptstyle \RR^4\times\CC^3,\ \mathrm{SL}(2)\times\mathrm{SL}(3)}\end{gathered}}} 
	\arrow[ur, bend left=12]  & & {\boxed{\begin{gathered}15\\[-0.7em] {\scriptstyle \RR^8\times\CC,\ \mathrm{Spin}(7)}\end{gathered}}} 
	\arrow[ul, bend right=10] & & & \\
	& & {\boxed{\begin{gathered}11\\[-0.7em] {\scriptstyle \CC^5,\ \mathrm{SL}(5)}\end{gathered}}} 
	\arrow[u] & & & & & \\
	& & & {\boxed{\begin{gathered}10\\[-0.7em] {\scriptstyle \CC^5,\ \mathrm{SL}(5)}\end{gathered}}} 
	\arrow[uur, bend right=20] 
	\arrow[ul, bend left=12] & & & & \\
	& & & * \arrow[u] & & & &
	\end{tikzcd}
	\caption{Orbit stratification of the type IIB nilpotence variety.} \label{table-iib}
\end{figure}

The orbit stratification is summarized in Figure~\ref{table-iib}. As above, we label orbits by their projective dimension together with the twisted background spacetime and the Levi of the stabilizer.

\section{Toward relating worldsheet and target space twists} \label{sec: app}
Given the above classification of twisting supercharges for type II supergravities, let us now discuss some applications. Here, we focus on three different areas: First, we show which twists of type IIA arise as dimensional reductions from eleven dimensions. Second, we discuss the worldsheet origins of some of the twists both for type IIA and IIB. Finally, we map out the relations between twists of IIA and IIB under T-duality.

\subsection{Dimensional reduction from eleven dimensions}
Type IIA supergravity is related to eleven-dimensional supergravity by compactification on a circle. Consequently, twists of type IIA supergravity can be obtained via dimensional reduction from twisted eleven-dimensional supergravity. 

Let $V_{11}$ denote the vector representation of $\Spin(11,\C)$ and $S_{11}$ the 32-dimensional spinor representation. The symmetric square of the spinor representation decomposes as $\Sym^2(S_{11}) \cong V_{11} \oplus \wedge^2 V_{11} \oplus \wedge^5 V_{11}$ and the supertranslation algebra for $\cN=1$ supersymmetry is given by
\begin{equation}
	\ft_{11d} = \Pi S_{11} \oplus V_{11}
\end{equation}
where the bracket between two odd elements is given by the projection $\gamma_{11d}: \Sym^2(S_{11}) \longrightarrow V_{11}$. As usual, the super Poincar\'e algebra is obtained as a semidirect product $\fp_{11d} = \mathfrak{so}(11,\C) \ltimes \ft_{11d}$.

To perform dimensional reduction to ten dimensions, we view $V = V_{\R} \otimes_{\R} \CC$ where $V_{\R}$ is equipped with a non-degenerate pairing. Then, we pick a real direction $v \in V$, along which we want to reduce. In particular, we have $\langle v,v\rangle_V > 0$ and we can write $V_{11} \cong V_{10} \oplus \C v$.
A dimensional reduction map along $v$ is a $\Spin(10)$-equivariant, surjective map of super Lie algebras $\pi : \ft_{11d} \longrightarrow \ft_{IIA}$ such that $\ker(\pi)= \C v$. A short computation using Schur's lemma shows that, up to scale, there are precisely two such maps. After identifying $S_{11} \cong S_+ \oplus S_-$ by restricting from $\Spin(11)$ to $\Spin(10)$, they are given on the odd elements as
\begin{equation}
	\pi_v|_{S_+} = \id_{S_+} \qquad \pi_v|_{S_-} = \id_{S_-} \qquad \text{and} \qquad \pi'_v|_{S_+} = \id_{S_+} \qquad \pi'_v|_{S_-} = -\id_{S_-}.
\end{equation}
Choosing one of these dimensional reduction maps is equivalent to specifying the chirality operator in ten dimensions in terms of the eleven-dimensional gamma matrix as $\pm \Gamma(v)$.

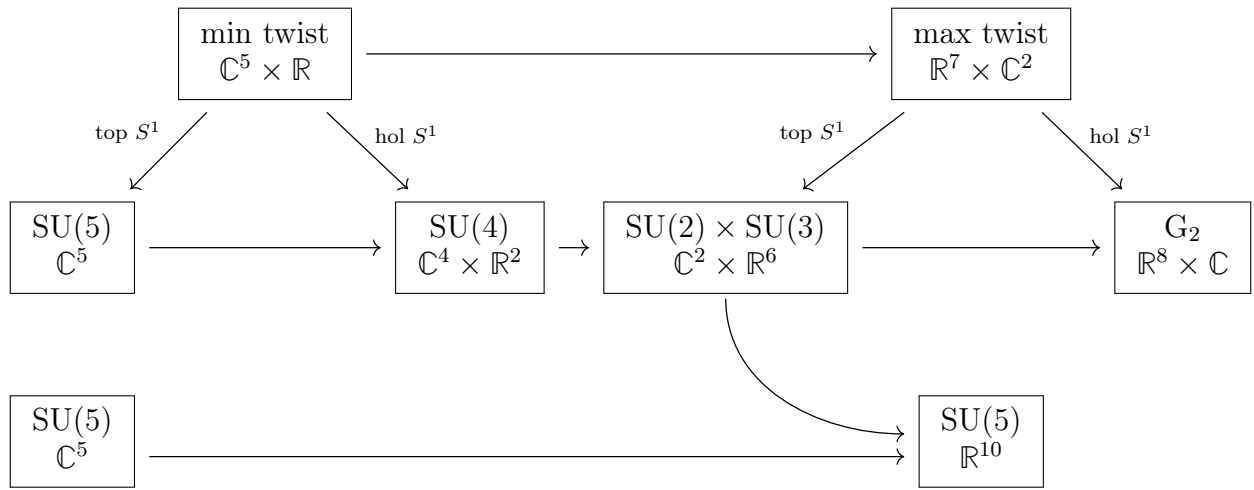
\begin{figure}
\begin{tikzcd}[row sep=2.5em, column sep=0.5em]
& 
\boxed{\begin{array}{c} \text{min twist} \\ \mathbb{C}^5 \times \mathbb{R} \end{array}} 
\arrow[dl, "\text{top } S^1"'] 
\arrow[dr, "\text{hol } S^1"] 
\arrow[rrrr] & 
& & & 
\boxed{\begin{array}{c} \text{max twist} \\ \mathbb{R}^7 \times \mathbb{C}^2 \end{array}} 
\arrow[dl, "\text{top } S^1"'] 
\arrow[dr, "\text{hol } S^1"] & \\
\boxed{\begin{array}{c} \mathrm{SU}(5) \\ \mathbb{C}^5 \end{array}} 
\arrow[rr] & & 
\boxed{\begin{array}{c} \mathrm{SU}(4) \\ \mathbb{C}^4 \times \mathbb{R}^2 \end{array}} 
\arrow[rr] & & 
\boxed{\begin{array}{c} \mathrm{SU}(2) \times \mathrm{SU}(3) \\ \mathbb{C}^2 \times \mathbb{R}^6 \end{array}} 
\arrow[rr] 
\arrow[dr, out=-90, in=180, yshift=0.1cm] & & 
\boxed{\begin{array}{c} \mathrm{G}_2 \\ \mathbb{R}^8 \times \mathbb{C} \end{array}} \\
\boxed{\begin{array}{c} \mathrm{SU}(5) \\ \mathbb{C}^5 \end{array}} \arrow[rrrrr, yshift=-0.2cm] & & & & & 
\boxed{\begin{array}{c} \mathrm{SU}(5) \\ \mathbb{R}^{10} \end{array}} &
\end{tikzcd}
\caption{Relations between twists in eleven dimensions and type IIA.}\label{fig: dim-red}
\end{figure}

In terms of the nilpotence varieties, a dimensional reduction map induces an inclusion
\begin{equation}
Y_{11d} \hookrightarrow Y_{IIA}. 
\end{equation}
Let us briefly summarize the structure of the nilpotence variety $Y_{11d}$. It is well known that $Y_{11d}$ decomposes into two orbits corresponding to the two different twists of eleven-dimensional supergravity.
\begin{itemize}
	\item[---] Maximal twisting supercharges form an orbit of dimension 22 and have stabilizer $(G_2 \times \mathrm{SL}(2)) \ltimes N_8$. The twisted theory has seven topological directions.
	\item[---] Minimal twisting supercharges form an orbit of dimension 15 and have stabilizer $\mathrm{SL}(5) \ltimes N_{15}$. The twisted theory has a single topological direction. 
\end{itemize}
These twists can both be reduced to type IIA by placing the $S^1$ either along a topological or a holomorphic direction. These compactifications were studied in~\cite{RSW}. The relation to the twists of type IIA is summarized in Figure~\ref{fig: dim-red}.
Here, the horizontal arrows denote further twists, while the vertical arrows indicate compactification on a circle that can either be in a holomorphic or in a topological direction.

At the same time, the diagram illustrates the interaction between the orbit decomposition and the dimensional reduction map. First, we see that the topological twist of type IIA cannot be obtained as a dimensional reduction from eleven dimensions. In other words, the corresponding twisting supercharges are neither in the image of $\pi_v|_{Y_{11d}}$ nor in that of $\pi'_v|_{Y_{11d}}$. 

The situation for the holomorphic twists is slightly different. Depending on the choice of dimensional reduction map, either the positive or negative chirality holomorphic twist is contained in the image, but never both at the same time.

In terms of the stratification of the IIA variety, we thus identify the top stratum of dimension 22 as the nilpotence variety $Y_{11d}$. The structure of $Y_{IIA}$ as a union between $Y_{11d}$ and a second component of projective dimension $21$ was already conjectured in~\cite{NV}.

\subsection{Toward relating worldsheet and target space twists}
Since supergravity theories are the low energy limits of ten-dimensional string theories, it is natural to expect that twisted supergravity theories can be obtained by modifying the worldsheet theory. Indeed, the conjectures on twisted supergravity by Costello and Li were informed by the A and B model topological string. A fully satisfactory relation between twists of the string worldsheet and target space twists of supergravity is, however, still lacking. Here, we take first steps in this direction, comparing the moduli spaces of twists for supergravity to those of the pure spinor superstring and linking the results to mixed A/B models.

\subsubsection{The pure spinor superstring}
Maybe the most promising formulation of string theory for the purpose of directly relating worldsheet and target space twists is the pure spinor superstring~\cite{Berkovits, Berkovits:2002zk}.

To introduce the pure spinor superstring in a flat background, we start with maps $\sigma: \Sigma \longrightarrow T$, where $\Sigma$ is a Riemann surface modeling the string worldsheet and $T = \RR^{1,9} \oplus \Pi S$ is viewed as an affine supermanifold serving as target space. For the type II string, $S$ is a 32-dimensional real spinor representation and one distinguishes two cases: For the IIA string, $S$ is the direct sum of two spinor representations of opposite chirality $S=S_+^\RR \oplus S_-^\RR$; for the IIB string it consists of two copies of the same chirality $S = S_+^{\RR} \oplus S_+^{\RR}$.

By choosing coordinates in target space, we expand a map $\sigma: \Sigma \longrightarrow T$ into components
\begin{equation}
\sigma(z,\bar{z}) = \left( X^\mu(z, \bar{z}), \theta_L^\alpha(z,\bar{z}), \theta_R^\beta(z, \bar{z}) \right),
\end{equation}
where $z$ and $\bar{z}$ coordinatize the worldsheet and $\mu=0,\dots,9$ is an index for the even directions of target space, while $\alpha,\beta = 1 \dots 16$ are indices for the odd directions. Further, we introduce conjugate momenta for the fermions that we denote by $p_{L \alpha}$ and $p_{R \beta}$.

To turn the pure spinor superstring into a consistent string theory, we supplement a ghost system
\begin{equation}
\lambda_L, \lambda_R: \Sigma \longrightarrow \cC \cP_\pm
\end{equation}
where $\cC \cP_\pm$ is the pure spinor cone and the chirality is determined depending on whether we are working with the IIA or the IIB string in the same way as for $\theta_{L/R}$ above. This means that $\lambda_L$ and $\lambda_R$ separately satisfy the constraint $\gamma(\lambda_L , \lambda_L) = 0 = \gamma(\lambda_R, \lambda_R)$.

The conjugate momenta for $(\lambda_L, \lambda_R)$ are denoted $(w_L, w_R)$ and are subject to the equivalence relation $w_{L/R \alpha} \sim w_{L/R \alpha} + \gamma^\mu_{\alpha \beta} \lambda_{L/R}^\beta$. Stated somewhat more abstractly, we can consider the map
\begin{equation}
\varphi_{L/R}: V \otimes \cO(\cC \cP_\pm) \otimes C^\infty(\Sigma) \longrightarrow S_{\pm} \otimes \cO(\cC \cP_\pm) \otimes C^\infty(\Sigma)  \qquad (\varphi_{L/R})^\mu_\alpha = \gamma^\mu_{\alpha \beta} \lambda_{L/R}^\beta
\end{equation}
where $\lambda^\alpha$ is now viewed as a generator in the ring of polynomial functions on the pure spinor cone. Then we have that $w_{L/R} \in \coker(\varphi_{L/R})$\footnote{Put more invariantly, $w_{L/R}$ are valued in Kähler differentials on the pure spinor variety.}.

The action for the pure spinor superstring is
\begin{equation}
S = \frac{1}{2 \pi \alpha'} \int \d z \d \bar{z} \left( \frac{1}{2} \partial X^\mu \bar{\partial} X_\mu + p_{L \alpha} \bar{\partial} \theta^\alpha_L + p_{R \alpha} \partial \theta^\alpha_R + w_{L \alpha} \bar{\partial} \lambda_L^\alpha + w_{R \alpha} \partial \lambda_R^\alpha \right).
\end{equation}
We note that the central charge of the matter sector is $10-32 = -22$, which is precisely compensated by the ghost system as $2 \dim(\cC \cP_\pm) = 2 \times 11 = 22$. The constraint on $(\lambda_L, \lambda_R)$ is very similar, but not identical to the square-zero conditions in the $\cN=(1,1)$ and $\cN=(2,0)$ super Poincar\'e algebras. In particular, the affine dimension of the top-dimensional component in the IIA and IIB nilpotence variety is 23.

The spectrum of the pure spinor superstring is obtained with respect to the BRST operator
\begin{equation}
Q_{BRST} = \int \d z \lambda_L ^\alpha d_{L \alpha} + \int \d \bar{z} \lambda_R^\alpha d_{R \alpha}
\end{equation}
where $d_{L \alpha} = p_{L \alpha} - \frac{1}{2} \gamma^\mu_{\alpha \beta} \theta^\beta_{L} \partial X^\mu - \frac{1}{8} (\gamma^\mu \theta_L)_\alpha (\theta_L \gamma_\mu \partial \theta_L)$ and an analogous formula holds for $d_{R \alpha}$. 

Thus, it is clear that assigning an expectation value to the pure spinor ghosts $(\lambda_L , \lambda_R)$ deforms the BRST operator and hence induces a twist. We therefore refer to the product of two pure spinor cones ($\cC \cP_+ \times \cC \cP_-$ for type IIA and $\cC \cP_+ \times \cC \cP_+$ for type IIB) as the moduli space of twists for the pure spinor superstring. This space is acted upon by $\Spin(10) \times \C^\times \times \C^\times$ where the two factors of $\C^\times$ are rescaling $\lambda_L$ and $\lambda_R$ separately.

As usual, we classify pairs $(\lambda_L, \lambda_R)$ by their intersection dimension. This gives the following results.

\subsubsection{Type IIA}
By Lemma~\ref{lem: ps}, the orbits of $\cC \cP_+ \times \cC \cP_-$ are classified by the following picture.
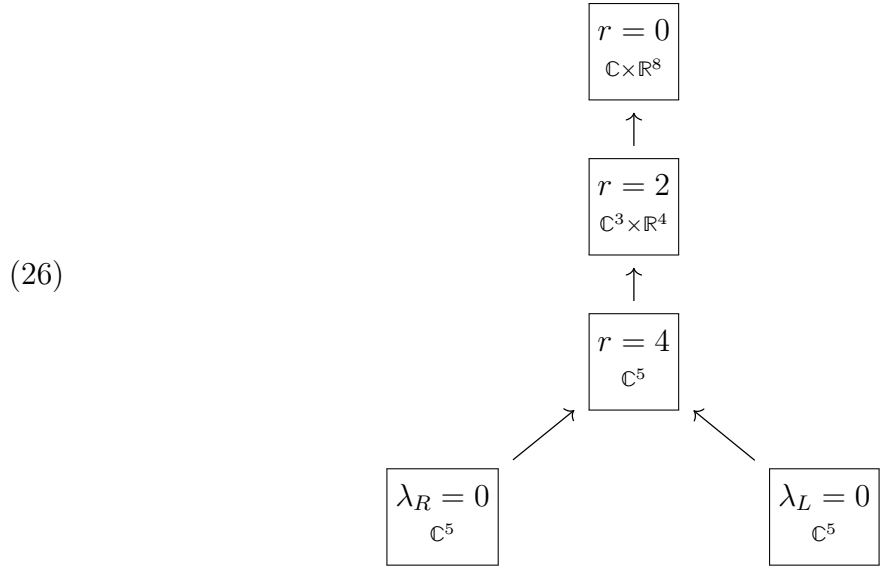
\begin{equation}
	\begin{tikzcd}[row sep=1.1em, column sep=2em, every node/.style={inner sep=3pt}]
	& {\boxed{\begin{gathered}r=0\\[-0.7em] {\scriptstyle \RR^{10}}\end{gathered}}} & \\
	& {\boxed{\begin{gathered}r=2\\[-0.7em] {\scriptstyle \mathbb{C}^2 \times \mathbb{R}^6}\end{gathered}}} \arrow[u] & \\
	& {\boxed{\begin{gathered}r=4\\[-0.7em] {\scriptstyle \mathbb{C}^4 \times \RR^2}\end{gathered}}} \arrow[u] & \\
	{\boxed{\begin{gathered}\lambda_R=0\\[-0.7em] {\scriptstyle \mathbb{C}^5}\end{gathered}}} \arrow[ur] & & {\boxed{\begin{gathered}\lambda_L=0\\[-0.7em] {\scriptstyle \mathbb{C}^5}\end{gathered}}} \arrow[ul]
	\end{tikzcd}
\end{equation}
Given a configuration $(\lambda_L, \lambda_R)$ in the moduli space of twists for the type IIA string, $Q= \lambda_L + \lambda_R$ is a square-zero supercharge in the type IIA super Poincar\'e algebra. Recall that the nilpotence variety $Y_{IIA}$ is the union between these two pure spinor cones and $Y_{11d}$. Thus, we obtain a map from the IIA pure spinor twists to those of IIA supergravity
\begin{equation}
	\cC \cP_+ \times \cC \cP_- \hookrightarrow Y_{IIA}
\end{equation}
that hits all twists except the orbits of $G_2$-equivariant twist that arise from the dimensional reduction of the maximal twist in eleven dimensions along a topological direction.

\subsubsection{Type IIB}
Given a configuration $(\lambda_L, \lambda_R)$ in the moduli space of twists for the type IIB string, we define a square-zero element in the type $(2,0)$ super Poincar\'e algebra as
\begin{equation}
Q = \lambda_L \otimes (1,0) + \lambda_R \otimes (0,1).
\end{equation}
To compare the moduli space of twists between the string theory and the supergravity theory, we first note that the orbit stratification between both spaces is slightly different. In the super Poincar\'e algebra, there is an $\mathrm{SO}(2,\C)$ R-symmetry, while the separation between left- and right-moving modes on the worldsheet breaks that symmetry. The classification of orbits in $\cC \cP_+ \times \cC \cP_+$ is given by the following diagram.
\begin{equation}
	\begin{tikzcd}[row sep=1.1em, column sep=2em, every node/.style={inner sep=3pt}]
	& {\boxed{\begin{gathered}r=1\\[-0.7em] {\scriptstyle \mathbb{C} \times \mathbb{R}^8}\end{gathered}}} & \\
	& {\boxed{\begin{gathered}r=3\\[-0.7em] {\scriptstyle \mathbb{C}^3 \times \mathbb{R}^4}\end{gathered}}} \arrow[u] & \\
	& {\boxed{\begin{gathered}r=5\\[-0.7em] {\scriptstyle \mathbb{C}^5}\end{gathered}}} \arrow[u] & \\
	{\boxed{\begin{gathered}\lambda_R=0\\[-0.7em] {\scriptstyle \mathbb{C}^5}\end{gathered}}} \arrow[ur] & & {\boxed{\begin{gathered}\lambda_L=0\\[-0.7em] {\scriptstyle \mathbb{C}^5}\end{gathered}}} \arrow[ul]
	\end{tikzcd}
\end{equation}
The mixed type twists for $r=1$ and $r=3$ match in a straightforward way with the corresponding twists in type IIB supergravity that were constructed from pairs of pure spinors. However, there are three different orbits of worldsheet twists that lead to a holomorphic dependence on target space as opposed to the two orbits that appear in the case of supergravity. Further, note that the $\Sp(4)$-equivariant twist is not in the image of the map $\cC \cP_+ \times \cC \cP_+ \hookrightarrow Y_{IIB}$.

\subsubsection{Mixed A/B models}
Starting from a worldsheet description of the superstring, it is expected that the theory behaves like a topological conformal field theory (TCFT) after performing a topological twist. The datum of a TCFT is equivalent to the choice of a Calabi--Yau $A_\infty$ category, which is interpreted as the category of branes~\cite{Costello:2004ei}. From here, one extracts descriptions of worldvolume theories on branes as derived endomorphisms and those of closed strings in target space by taking cyclic cochains of the brane category. The former is related to twists of supersymmetric gauge theories~\cite{Yoo:2025qlw}, while the latter theories are expected to match twisted supergravities~\cite{CostelloLi}.

Twists of worldsheet-like models have been widely studied in the $\cN=(2,2)$ sigma model, which admits two types of topological twists, the A- and the B-model~\cite{Witten:1991zz}. The categories of A- and B-branes are described by the Fukaya category and the category of coherent sheaves on target space, respectively. The target space theory of the B-model is BCOV theory~\cite{BCOV} which is holomorphic, while the A-model gives rise to a topological theory whose BV complex is described by de Rham forms on target space.\footnote{Note that this description holds in perturbation theory; the full A-model receives important non-perturbative corrections from instantons.}

Mixed A/B models can be constructed when the target space has a product structure $X\times Y$ and we can consider the product category $\mathrm{Br}_A(X) \otimes \mathrm{Br}_B(Y)$, which combines A-branes on $X$ with B-branes on $Y$. Working in a flat background, this gives rise to mixed topological-holomorphic theories in target space with BV complex
\begin{equation}
	(\Omega^\bu(\R^{10-2r}) , \d_{dR}) \otimes \mathrm{BCOV}(\C^{r}).
\end{equation}
In such a model, branes have worldvolumes of real dimensions $5-r + 2k$ for $k \in \{0, \dots , r\}$. Since type IIA features odd-dimensional and type IIB even-dimensional branes, we thus expect mixed A/B models with even $r$ to be realized as twists of type IIA supergravity, while those with odd $r$ are related to twists of type IIB supergravity.

This matches well with our findings on the moduli space of twists in supergravity: Constructing a target space twist from a pair of pure spinors, we obtain a twisting supercharge where the number of holomorphic directions is given by the intersection dimension $r$ of their associated isotropic subspaces. These are the candidate target space twists realizing mixed A/B models. Indeed, we found that $r$ is even for type IIA and odd for type IIB. Further, the semi-simple piece of the stabilizer of such target space twisting supercharges is a product $\SU(5-r) \times \SU(r)$, consistent with the expectation that a mixed A/B model can be defined on products of Calabi--Yau manifolds. A special case occurs for $r=1$: Here, the B-model direction is deformed by a linear superpotential that renders the twisted theory locally trivial in field space. One way to see this is to view the twist as a further deformation of BCOV theory on $\CC^5$.

Finally, there are twists of supergravity that are not mixed A/B models. For type IIA, these are the two holomorphic twists and the $G_2$ twist; for type IIB, the $\Spin(7)$- and $\Sp(4)$-equivariant twists. These come in two types. First, it is natural to expect that such twists with special holonomy groups as their stabilizers are related to topological worldsheet theories described by categories of branes in special holonomy manifolds~\cite{TopoM}. On the other hand, the $\SU(5)$-twists in type IIA are most likely not obtained from a topological twist of a worldsheet theory. Indeed, in the pure spinor string, such twists are obtained by only putting an expectation value on either the left- or right-moving modes, which might lead to a holomorphic worldsheet theory instead of a topological one.

\subsection{T-duality}
T-duality connects type IIA and type IIB string theory, thus it is natural to ask for the relation between the different twisting supercharges. This is most directly seen for target space twisting supercharges built from a pair of pure spinors, or, equivalently, in terms of twists on the pure spinor worldsheet. We start by discussing these cases and then turn our attention to supercharges containing impure spinors below.

Let $v \in \R^{10}$ be the direction where we would like to place the circle. Then T-duality acts on the pure spinor ghosts by Clifford multiplication on the right-moving ghost~\cite{Benichou:2008it}, i.e.
\begin{equation}
	(\lambda_L, \lambda_R) \mapsto (\lambda_L, v \cdot \lambda_R)
\end{equation}
where we now view $v$ as an element of the complex vector representation $V$. From here, it is straightforward to see how the different orbits are related under this map.
\begin{itemize}
	\item[---] The twists with $\lambda_L=0$ or $\lambda_R=0$ are mapped to one another.
	\item[---] For mixed type twists labeled by the intersection dimension $r$, applying T-duality along a topological direction increases the intersection dimension to $r+1$, along a holomorphic direction the intersection dimension to $r-1$.
\end{itemize}
This is consistent with the expectation that in mixed A/B models, T-duality exchanges topological and holomorphic directions.

This discussion immediately applies to target space twists that are constructed from a pair of pure spinors. However, for those target space twists that do not arise from the pure spinor worldsheet, T-duality does not work in such a straightforward way. For example, applying the map from type IIA to type IIB supercharges,
\begin{equation}
	T_v: S_+ \oplus S_- \longrightarrow S_+ \otimes \C^2 \qquad (\psi_+, \psi_-) \mapsto \psi_+ \otimes (1,0) + (v \cdot \psi_-) \otimes (0,1),
\end{equation}
we see that this map in general does not preserve the square-zero condition. Indeed, a short calculation shows
\begin{equation}
	[T_v(Q), T_v(Q)]_{IIB} = (1- \langle v,v \rangle_V) f + 2 \langle v, f \rangle_V v
\end{equation}
where $f = \gamma(\psi_+, \psi_+) = - \gamma(\psi_-, \psi_-)$. Thus, for $Q$ in the $G_2$ orbit of the IIA nilpotence variety, $T_v(Q)$ is only square-zero when $\langle v,v\rangle_V = 1$ and $\langle v, f \rangle_V = 0$.

In other words, the map $T_v$ always restricts to a map on the products of pure spinor cones in which the pure spinor worldsheet ghosts take values; however, it does not restrict to a map between the nilpotence varieties $Y_{IIA}$ and $Y_{IIB}$, as indicated in the following diagram. 
\begin{equation}
	\begin{tikzcd}[row sep=large, column sep=huge]
	S_+ \oplus S_- \arrow[r, "T_v"] & S_+ \oplus S_+ \\
	Y_{IIA} \arrow[u, hook] \arrow[r, dashed, "\diagup \diagup" description] & Y_{IIB} \arrow[u, hook'] \\
	\cC \cP_+ \times \cC \cP_- \arrow[u, hook] \arrow[r, "T_v|_{\cC \cP_+ \times \cC \cP_-}"] & \cC \cP_+ \times \cC \cP_+ \arrow[u, hook']
	\end{tikzcd}
\end{equation}

In the following, let us investigate in more detail how $T_v$ acts on $G_2$-twisting supercharges whenever the result is square-zero.

Recall that for the $G_2$-twist, the vector representation splits into one complex direction spanned by $(f, f^\vee)$ and eight topological directions. We choose $v$ aligned with the topological directions and fix it to be of unit norm so that the square-zero condition is preserved. In terms of $G_2$ representations, these eight directions decompose as $1_{G_2} \oplus V_7$. Thus, we distinguish between two cases: We can place the circle on the direction that is stabilized by $G_2$ or we break the full $G_2$-equivariance by picking a direction $v \in V_7$.

First, let $v \in 1_{G_2}$. Then we find that $v \cdot \psi_-$ is a $G_2$-invariant spinor in $S_+$. Recall that the spinor representation decomposes under $G_2$ as
\begin{equation}
	S_+ = (1_{G_2} \oplus V_7) \oplus (1_{G_2} \oplus V_7) \otimes \langle f^\vee \rangle .
\end{equation}
Note that $f \cdot \psi_\pm$ implies that $f \cdot v \cdot \psi_- = 0$ via the Clifford algebra and hence $v \cdot \psi_-$ lands in the former summand. By $G_2$-equivariance, we thus have $v \cdot \psi_- = c \psi_+$ for some $c \in \C$. Finally, by applying Clifford multiplication with $v$ twice, we find that $c= \pm i$. In summary, we therefore have that under T-duality
\begin{equation}
	Q = \psi_+ + \psi_- \mapsto \psi_+ \otimes (1, \pm i),
\end{equation}
where we recognize the right hand side as a $\Spin(7)$-equivariant twisting supercharge of type IIB. This means we can consider both the $G_2$-twist of IIA and the $\Spin(7)$-twist of IIB on manifolds of the form $M_7 \times S^1 \times \Sigma$, where $M_7$ is a $G_2$ manifold and $\Sigma$ is a complex manifold of dimension one. Performing T-duality on the $S^1$ then relates both twists.  

Second, we can break $G_2$-equivariance by taking $v \in V_7$. In that case, we obtain a type IIB twisting supercharge
\begin{equation}
	\psi_+ \otimes (1,0) + (v \cdot \psi_-) \otimes (0,1)
\end{equation}
and since $\psi_+$ and $v \cdot \psi_-$ are linearly independent, the supercharge is of rank 2. To work out in which of the rank 2 orbits this supercharge lands, we study the intersection pattern between the pure spinor variety and the span of $\psi_+$ and $v \cdot \psi_-$. Note that
\begin{equation}
	\gamma(a \psi_+ + b (v \cdot \psi_-), a \psi_+ + b (v \cdot \psi_-) ) = (a^2 - b^2) f + 2 a b \gamma(\psi_+, v \cdot \psi_-).
\end{equation}
We now show that the mixed term vanishes. First, we verify this for an explicit representative. For example, with the $G_2$-twisting supercharge from Table~\ref{table-iia}, we have $\psi_+ = 1 + e_{2345}^\vee$ and $\psi_-= e_2^\vee - e_{345}^\vee$ and can take $v = e_2 + e_2^\vee$. Then, $v \cdot \psi_- = 1 - e_{2345}^\vee$, so that the mixed term vanishes. Acting with $G_2$ on $v$, it follows that $\gamma(\psi_+ , v \cdot \psi_-) = 0$ for all $v \in V_7$. 

Any other twisting supercharge in the $G_2$-orbit is obtained by acting with some $g \in \Spin(10)$ on $\psi_+$ and $\psi_-$. Note that for $\psi'_\pm = g(\psi_\pm)$, the subspace $V_7$ is also transformed to $V_7' = g(V_7)$ so that all vectors $v' \in V_7$ can be written as $v' = g(v)$ for some $v \in V_7$. Thus, $\gamma(\psi'_+ , v' \cdot \psi'_-) = \gamma(g(\psi_+) , g(v) \cdot g(\psi_-) ) = 0$ by $\Spin(10)$-equivariance.

Thus, there are always two lines of pure spinors in the span corresponding to $a = \pm b$. Using the classification pattern of~\cite{iib-nv}, we therefore identify the T-dual twist as the $\mathrm{SU}(4)$-twist of type IIB. In this setting, we can consider both theories on $(X_3 \times S^1) \times S^1 \times \Sigma$, where $X_3$ is a Calabi--Yau three-fold and we either consider $X_3 \times S^1$ as a $G_2$-manifold or $(X_3 \times S^1) \times S^1$ as a Calabi--Yau four-fold. As before, $\Sigma$ is a complex manifold of dimension one on which the theory depends holomorphically. T-duality along the first circle relates both twisted theories.

\subsection{Outlook}
Our findings on the type IIA nilpotence variety raise the following question. We found $Y_{IIA}$ to be a union of two components $Y_{IIA} = Y_{11d} \cup C$, with the component $C$ being identified with the moduli space of twists for the pure spinor superstring and $Y_{11d}$ those twists of IIA obtained from dimensional reduction. Note that these two components have a large intersection so that most twists admit both a pure spinor worldsheet and an eleven-dimensional description. The $G_2$-invariant twist of type IIA, however, cannot be obtained by giving a expectation values to the pure spinor ghosts on the worldsheet. Thus, it is natural to ask what operation on the worldsheet recovers this twist in the supergravity theory.

A more systematic comparison between worldsheet and target space twists is complicated by the fact that the notion of twisting the string worldsheet is dependent on the framework used for the worldsheet theory. Here we assigned expectation values to the pure spinor worldsheet ghosts, but a satisfactory treatment would employ a definition that more clearly parallels Costello and Li's definition of twiste supergravity in target space. One way to do this is to work with the RNS string before gauge fixing, viewing it as a two-dimensional theory coupled to worldsheet conformal supergravity. Supersymmetry is then gauged on the worldsheet, and the definition of~\cite{CostelloLi} applies directly. We hope to return to this elsewhere.

The results also suggest computations in the language of brane categories. For every twisted supergravity theory originating from a topological worldsheet an associated Calabi--Yau $A_\infty$ category is expected. The BV fields of the twisted supergravity theory are then obtained by taking cyclic cochains of this category. For the mixed type twists these categories are products of A and B branes; extracting such a category from a twist of a type II string---be it in the RNS, GS, or pure spinor description---would close the loop and establish a firm correspondence between worldsheet and target space twists. Further, even if the special holonomy twists are not realized as expectation values in the pure spinor ghosts, they can still have an underlying description by a topological worldsheet. If so, then they are associated with categories of branes on manifolds of $G_2$-, $\Spin(7)$-, and $\Sp(4)$-holonomy.

\printbibliography

\end{document}